\def\ref{\par\noindent\hang}

\def\spose#1{\hbox to 0pt{#1\hss}}
\def\approxlt{\mathrel{\spose{\lower 3pt\hbox{$\sim$}}
	\raise 2.0pt\hbox{$<$}}}
\def\approxgt{\mathrel{\spose{\lower 3pt\hbox{$\sim$}}
	\raise 2.0pt\hbox{$>$}}}

\def\multleft#1{\hbox to size{\vbox {\halign {\lft{##}\cr #1}}\hfill}\par}
\def\multright#1{\hbox to size{\vbox {\halign {\rt{##}\cr #1}}\hfill}\par}

\def\degmark{^\circ}
\def\$<${\thinspace}
\def\s{\hbox{\phantom{5}}}	

\def\boxit#1{\vbox{\hrule\hbox{\vrule\kern3pt\vbox{\kern3pt
          #1 \kern3pt}\kern3pt\vrule}\hrule}}

\def\cm{{\rm\thinspace cm}}

\def\erg{{\rm\thinspace erg}}

\def\keV{{\rm\thinspace keV}}

\def\Msun{\hbox{$\rm\thinspace M_{\odot}$}}

\def\s{{\rm\thinspace s}}

\def\cmsq{\hbox{$\cm^2\,$}}

\def\ps{\hbox{$\s^{-1}\,$}}

\documentstyle[psfig,12pt]{article}

\title{Iron fluorescence from within the innermost stable orbit of black hole
accretion disks}

\author{Christopher~S.~Reynolds\thanks{email:chris@rocinante.colorado.edu} and Mitchell~C.~Begelman\thanks{Also at
Department of Astrophysical, Planetary, and Atmospheric Sciences,
University of Colorado, Boulder, Colorado.}\\
{\small JILA, University of Colorado, Boulder, Colorado, CO~80309-0440}
}

\begin{document}

\maketitle

\begin{abstract}
The fluorescent iron K$\alpha$ line is a powerful observational probe of
the inner regions of black holes accretion disks.  Previous studies
have assumed that only material outside the radius of marginal stability
($r=6m$ for a Schwarzschild hole) can contribute to the observed line
emission.  Here, we show that fluorescence by material {\it inside} the
radius of marginal stability, which is in the process of spiralling towards
the event horizon, can have a observable influence on the iron line profile
and equivalent width.  For concreteness, we consider the case of a
geometrically thin accretion disk, around a Schwarzschild black hole, in
which fluorescence is excited by an X-ray source placed at some height
above the disk and on the axis of the disk.  Fully relativistic line
profiles are presented for various source heights and efficiencies.  It is
found that the extra line flux generally emerges in the extreme red wing of
the iron line, due to the large gravitational redshift experienced by
photons from the region within the radius of marginal stability.

We apply our models to the variable iron line seen in the {\it ASCA}
spectrum of the Seyfert nucleus MCG$-$6-30-15.  It is found that the change
in the line profile, equivalent width, and continuum normalization, can be
well explained as being due to a change in the height of the source above
the disk.  Thus, we can explain the iron line properties of MCG$-$6-30-15
within the context of an accretion disk around a non-rotating black hole.
This contrasts with previous studies which, due to the absence of
fluorescence from within the radius of marginal stability, have indicated
that this object possesses a rapidly rotating (i.e., near-extremal Kerr)
black hole.  This is an important issue since it has direct bearing on the
spin paradigm for the radio-loud/radio-quiet dichotomy seen in accreting
black hole systems.  We discuss some future observational tests which could
help distinguish slowly rotating black holes from rapidly rotating holes.
\end{abstract}

\section{Introduction}

The fluorescent K$\alpha$ emission line of cold\footnote{By `cold', we mean
in the ionization range Fe\,{\sc i}--Fe\,{\sc xvii}} iron provides a unique
diagnostic of the inner regions of accretion flows around black holes.
Such fluorescent lines are produced when regions of optically-thick, cold
material are externally illuminated by hard X-rays (George \& Fabian 1991;
Matt, Perola \& Piro 1991).  In particular, illumination of a
standard thin accretion disk by X-rays from a disk-corona will produce
fluorescent iron line emission.  Since the line energy is well known
($6.40\keV$) and its intrinsic width is very small, observations of the
line profile can provide direct information on the Doppler shifts and
gravitational redshifts affecting the line-emitting material (Fabian et
al. 1989; Laor 1991).

The {\it Advanced Satellite for Cosmology and Astrophysics} ({\it ASCA})
affords us the X-ray spectral capability to perform such an
investigation.  It has been found that the iron lines in many Seyfert 1
nuclei, previously discovered by {\it Ginga} (Pounds et al. 1990), possess
the profile expected if they were to originate from the inner regions of a
thin accretion disk around a (Schwarzschild or Kerr) black hole.  The large
widths and skewnesses of these lines are well explained as being due to a
combination of Doppler shifts and gravitational redshifts (Mushotzky et
al. 1995; Tanaka et al. 1995; Nandra et al. 1997).  In the highest
signal-to-noise examples (e.g., the Seyfert 1 nucleus MCG$-$6-30-15, Tanaka
et al. 1995), alternative models for producing a broad, skewed line can be
examined and rejected (Fabian et al. 1995).

Given this interpretation, broad iron lines can be used to study the
astrophysics of accretion within the immediate vicinity of a black hole.
There is still much to be learnt.  Firstly, many active galactic nuclei
(AGN) display iron lines that are significantly stronger than predicted by
the standard X-ray reflection model.  This may be providing evidence of
significantly non-solar abundances in the accretion flow (George \& Fabian
1991; Reynolds, Fabian \& Inoue 1996), or the enhancement of the line
strength via relativistic effects (Martocchia \& Matt 1996; Reynolds \&
Fabian 1997).  Secondly, there are hints that higher-luminosity AGN tend to
show more highly ionized iron lines or no iron lines at all (Nandra et
al. 1995, 1996).  Similar results are seen for Galactic Black Hole
Candidates (GBHC).  This may suggest that these objects possess more highly
ionized accretion disks, as expected if they are operating closer to the
Eddington limit.  Thirdly, some AGN display extremely strong lines with
very broad red wings which cannot be fit with the standard model of iron
fluorescence about a Schwarzschild hole (Iwasawa et al 1996, hereafter I96;
Reynolds 1997).  In these cases, it has been suggested that the line
emission originates from disks around near-extremal Kerr holes (Dabrowski
et al. 1997).

General relativity predicts the existence of a critical radius within which
massive particles cannot occupy stable circular orbits (see, e.g., Misner,
Thorne \& Wheeler 1973).  We shall denote this {\it radius of marginal
stability} as $r_{\rm ms}$.  For a Schwarzschild black hole, $r_{\rm
ms}=6m$, where we have set $c=G=1$.  Standard models of iron fluorescence
from around black holes assume that there is no fluorescence (and almost no
material!) within $r_{\rm ms}$.  For rotating (Kerr) holes, $r_{\rm ms}<6m$
and decreases monotonically as the dimensionless spin-parameter of the hole,
$a$, increases ($r_{\rm ms}\rightarrow m$ as $a\rightarrow 1$).  Under the
assumption that all iron fluorescence occurs outside $r_{\rm ms}$, this is
the basis for the statement that very broad red-wings imply a Kerr geometry:
near-extremal Kerr geometry is required in order to make $r_{\rm ms}$
sufficiently small, and hence the relativistic effects sufficiently strong,
to explain such lines.  

In this paper, we suggest that iron fluorescence from material within
$r=r_{\rm ms}$ may, indeed, be observable.  In particular, this emission
may be relevant to the extremely broad and strong lines mentioned above.
In Section 2, we discuss the simplified disk model that we assume for our
calculations.  We show that, for typical parameters, the optical depth of
the disk is sufficient for fluorescence even far within $r=r_{\rm ms}$.
Section 3 describes our disk-line calculations, which include fluorescence
from within $r=r_{\rm ms}$, and presents results for various regions of
parameter space.  Section 4 discusses the relevance of these calculations
to existing observational issues and possible future tests of this model.
Our conclusions are drawn in Section 5.

\section{Innermost regions of a black hole accretion disk}

We consider an accretion disk around a Schwarzschild black hole of mass
$m$.  It is assumed that the accretion disk is geometrically thin and
radiatively efficient as opposed to, for example, a geometrically thick
advection dominated disk.  It is also assumed that the disk is illuminated
by some nearby hard X-ray source, such as the high-energy emission from a
disk-corona.  This illumination will drive iron line fluorescence.  We
suppose that the X-ray illumination is axisymmetric and denote by $g(r)$
the illuminating flux per unit area of the disk.  We shall refer to $g(r)$
as the {\it illumination law}.  Note that for matter at a given ionization
state, the iron fluorescent emission will be proportional to $g(r)$.

There is a large body of theoretical work on the global structure of thin
accretion disks around black holes (e.g. Muchotrzeb \& Pacynski 1982;
Abramowicz \& Kato 1989; Chen \& Taam 1993).  These studies show that the disk has
the following properties.  At radii $r\approxgt r_{\rm ms}$, the disk material is
essentially in circular, Keplerian motion superposed with a small
radial inflow.  This continues down to $r\sim r_{\rm ms}$.  Within
$r=r_{\rm ms}$ the material starts to spiral towards the hole.  It soon
passes through a sonic point (located near $r=r_{\rm ms}$).  There
is very little dissipation once the material has passed the sonic point:
the subsequent motion approximately conserves energy and angular momentum.

\subsection{Velocity field}

On the basis of the studies discussed in the previous paragraph, we
approximate the full disk velocity field by splitting the disk into two
regions:
\begin{enumerate}
\item $r>r_{\rm ms}=6m$ -- the material is assumed to be in perfect Keplerian
motion.  In Schwarzschild co-ordinates, this gives the following components
for the 4-velocity vector:
\begin{eqnarray}
u^t& \equiv & \dot{t} = \frac{r}{\sqrt{r(r-3m)}}\\
u^\phi &\equiv & \dot{\phi} = \frac{1}{r}\sqrt{\frac{m}{r-3m}}\\
u^r &\equiv &\dot{r} = 0\\
u^\theta &\equiv &\dot{\theta} = 0
\end{eqnarray}
where the dot represents differentiation by the material's proper time.
\item $r<r_{\rm ms}=6m$ -- the material is assumed to be in free-fall with the
energy and angular momentum of material at the innermost stable orbit.
This gives the following components of the 4-velocity vector:
\begin{eqnarray}
u^t &=&\frac{r\sqrt{8}}{3(r-2m)}\\
u^\phi &=&\frac{2\sqrt{3}m}{r^2}\\
u^r &=&-\sqrt{\frac{8}{9}-\left(1-\frac{2m}{r}\right)\left(1+\frac{12m^2}{r^2}\right)}\\
u^\theta &=& 0
\end{eqnarray}
where we have used the fact that the specific angular momentum of material
at the innermost stable orbit is $l=2\sqrt{3}m$ and its specific energy is
$E=\sqrt{8/9}$.
\end{enumerate}
The generalization of these formulae to the case of Kerr geometry is
straightforward, but shall not be performed here.

\subsection{Optical depth}

\begin{figure}
\hbox{
\hspace{1cm}
\centerline{\psfig{figure=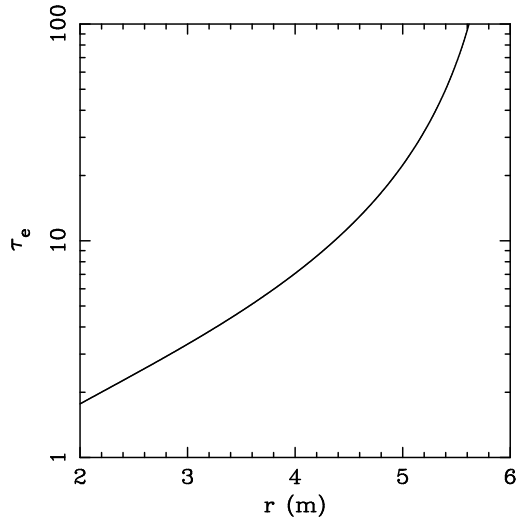,width=0.65\textwidth,angle=270}}
}
\caption{Electron scattering optical depth $\tau_{\rm e}$ as a function of
radius for a Schwarzschild hole with $L=0.1L_{\rm Edd}$ and $\eta=0.06$.
See text for the approximations used.  The dependences of $\tau_{\rm e}$ on
$L/L_{\rm Edd}$ and $\eta$ are given in eqn (11).}
\end{figure}

The disk must remain Thomson thick and relatively cold in order to produce
a significant fluorescent iron line upon hard X-ray illumination.  The
optical depth of the disk can be calculated from the (baryon number)
continuity equation,
\begin{equation}
(\rho u^\mu)_{;\mu}=0,
\end{equation}
where $\rho$ is the mass density in the comoving frame.  For a
steady-state, axisymmetric, thin disk, this reduces to
\begin{equation}
r\Sigma u^r = -\frac{\dot{m}}{2\pi},
\end{equation}
where $\Sigma$ is the (comoving) surface mass density and $\dot{m}$ is the
accretion rate.  Since the electron scattering optical depth normally
through the disk in the comoving frame is given by $\tau_{\rm e}=\Sigma
\sigma_{\rm T}/m_{\rm p}$, we can cast the continuity equation in the form,
\begin{equation}
\tau_{\rm e}=\frac{2m}{\eta r (-u^r)}\frac{L}{L_{\rm Edd}},
\end{equation}
where $L_{\rm Edd}$ is the Eddington luminosity of the black hole and
$\eta$ is the overall radiative-efficiency ($\eta\approx 0.06$ for a
Schwarzschild hole).  Figure~1 plots $\tau_{\rm e}$ as a function of $r$
($r<r_{\rm ms}$) for the case where $L=0.1L_{\rm Edd}$ and $\eta=0.06$.  It
can be seen that the disk remains optically-thick almost all of the way
down to the horizon of the hole.  Note that the free-fall approximation
becomes progressively better as one considers smaller $r$.  Thus, this
result should remain valid even when one considers fully self-consistent,
global disk models.

\subsection{X-ray illumination}

The largest uncertainty in the study of fluorescent iron lines from
accretion disks is the illumination law $g(r)$.  Indeed, one might
eventually hope to use well-studied iron emission lines to determine $g(r)$
and thus learn the geometry of the X-ray source.  For the present purposes,
we are interested in illumination laws that lead to appreciable
illumination of the region $r<r_{\rm ms}$.  The X-ray flux from a
geometrically thin disk-corona most likely reflects the local energy input
which is, presumably, related to the local viscous dissipation in the
accretion disk.  Since there is very little dissipation in the region
$r<r_{\rm ms}$, a thin corona is most likely {\it not} an appropriate
geometry if we wish to explore the possibility of iron lines from $r<r_{\rm
ms}$.  This geometry has been adopted by Dabrowski et al. (1997) and the
resulting iron line profiles have been explored using both Schwarzschild
and Kerr metrics.

Here, we adopt the following idealized X-ray source geometry: the source is
assumed to be an isotropic point source situated on the rotation axis of
the accretion disk at a height $h$ above the disk plane.  This may be
viewed as an approximation to any source geometry which is axisymmetric and
significantly away from the disk plane (for example, a quasi-spherical
disk-corona, or X-ray emission from a jet).  The source spectrum is assumed
to be a power-law with energy index $\alpha=1$ (emitted power
$F(\nu)\propto \nu^{-\alpha}$) for all frequencies of interest.

\begin{figure}
\hbox{
\hspace{1cm}
\centerline{\psfig{figure=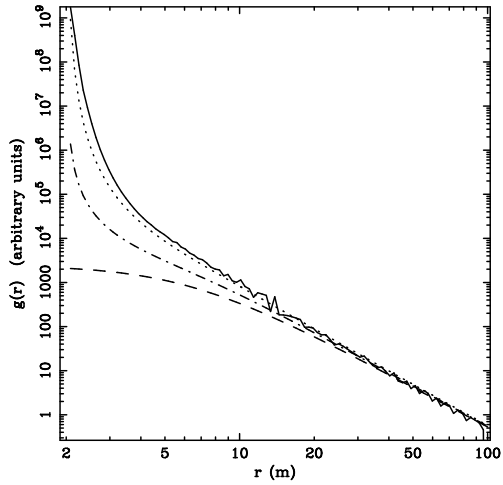,width=0.65\textwidth,angle=270}}
}
\caption{Illumination law for the case of a point source at a distance
$h=6m$ above the center of the accretion disk.  The source is assumed to be
isotropic and have an energy index of $\alpha=1$.  The solid line shows the
full illumination law, $g(r)$, including all relativistic effects.  The
dashed line shows the Euclidean result ($g_1(r)$ in the main text).  The
dot-dashed line shows the result of neglecting Doppler corrections
($g_2(r)$ in the main text).  The dotted line shows the function $g_3(r)$,
which acts as a good approximation to $g(r)$.}
\end{figure}

In the absence of any relativistic effects, the resulting illumination law
is given by
\begin{equation}
g_1(r)\propto (r^2 + h^2)^{-3/2}.
\end{equation}
Gravitational and Doppler shifts will have a significant influence on this
illumination law.  In the unphysical case of a static disk, only
gravitational shifts are important and the resulting relativistic
modifications to eqn (12) are
\begin{equation}
g_2(r)\propto \frac{g_1(r)}{(1-2m/r)^{(3+\alpha)/2}},
\end{equation}
where we have used the phase-space invariant $I_\nu/\nu^3$ and included the
gravitational `k-correction'.  For small $r$ this gravitational term is of
great importance and can enhance the illumination of the inner disk by
orders of magnitude.

In reality, the disk material is moving and so there are Doppler
corrections.  To calculate the correct illumination law, including all
relativistic effects, we used a photon tracing code (described in Section
3) to simulate the isotropic emission of photons from the source.  For
those photons that were found to strike the disk-plane (defined as
$\theta=\pi/2$), we calculate the gravitational/Doppler redshift from
\begin{equation}
1+z=\frac{(u^\mu p_\mu)_{\rm source}}{(u^\mu p_\mu)_{\rm disk}},
\end{equation}
where $u^\mu$ is the 4-velocity of the source/disk material and $p_\mu$ is
the contravariant form of the photon's 4-momentum at the source/disk.  The
resulting enhancement in the illuminating intensity at a given frequency is
$(1+z)^{-3-\alpha}$.  Adding up the effects of many simulated photons, we
can construct the resulting illumination law $g(r)$.  This is shown in
Fig.~2 for the case of $h=6m$, along with $g_1(r)$ and $g_2(r)$ for
comparison.  Figure~2 also shows $g_3(r)$, defined by
\begin{equation}
g_3(r)\propto\frac{g_1(r)}{(1-2m/r)^{3+\alpha}}\propto\frac{g_2(r)}{(1-2m/r)^{(3+\alpha)/2}}.
\end{equation}
It can be seen that $g_3(r)$ is a good approximation to the full
(numerical) illumination law $g(r)$ for all but the very innermost radii.
This amounts to saying that the Doppler effect and pure gravitational
redshift effects are of approximately equal importance.  In the
calculations presented in this paper, we shall use $g_3(r)$ as a convenient
approximation to the full illumination law.

\subsection{Ionization of the inner disk}

As shown above, relativistic effects can lead to a large enhancement in the
irradiation of the inner regions of the disk.  Since we are primarily
interested in the resulting iron fluorescence, we must address the issue of
the ionization state of the matter.

\begin{figure}
\hbox{
\hspace{1cm}
\centerline{\psfig{figure=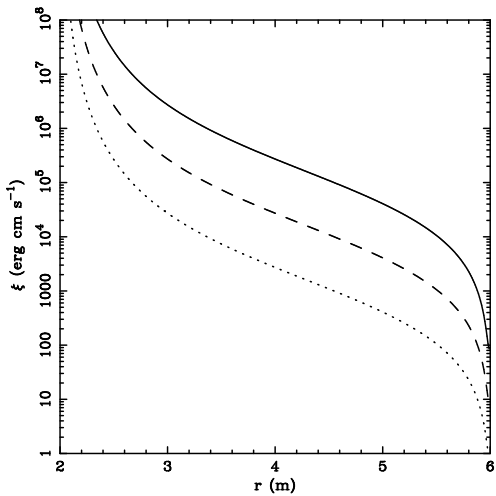,width=0.65\textwidth,angle=270}}
}
\caption{Ionization parameter $\xi$ as a function of radius for $h=6m$ and
$\eta_{\rm x}=0.06$ (solid line), $\eta_{\rm x}=0.006$ (dashed line) and
$\eta_{\rm x}=0.0006$ (dotted line).}
\end{figure}

We are concerned with accretion disks that are strongly irradiated by an
external X-ray source.  Thus, photoionization by the external irradiating
flux is likely to be the dominant process determining the ionization state
of the disk material.  Photoionization of the region $r>r_{\rm ms}$ has
been considered by Matt, Fabian \& Ross (1993, 1996), who find that cold
iron fluorescence results provided that $L\approxlt 0.1L_{\rm Edd}$.
Similar calculations have not been performed explicitly for the region
$r<r_{\rm ms}$.  Here we adopt a simple, but useful, approach to this
problem.  We define the X-ray ionization parameter, $\xi$, as
\begin{equation}
\xi (r)=\frac{4\pi F_{\rm X}(r)}{n(r)},
\end{equation}
where $F_{\rm X}(r)$ is the X-ray flux (defined over some fixed energy
band) striking a unit area of the disk at radius $r$, and $n(r)$ is the
comoving electron number density.  Now, using that fact that $g_3(r)$
approximates the full illumination law, we have
\begin{equation}
F_{\rm X}\approx \left(\frac{L_{\rm X}h}{4\pi (r^2+h^2)^{3/2}}\right)\left(\frac{1-2m/h}{1-2m/r}\right)^{3+\alpha}.
\end{equation}
In this expression, the first bracket on the right hand side is the
standard Euclidean result, whereas the second bracket is the (approximate)
relativistic correction.  The average density of the disk material is
easily found from the continuity equation (10) to be
\begin{equation}
n=\frac{\dot{m}}{4\pi r h_{\rm disk} (-u_r) m_{\rm p}},
\end{equation}
where $h_{\rm disk}$ is the half thickness of the disk.  If we define the
X-ray efficiency as $L_{\rm X}=\eta_{\rm x}\dot{m}$, then the ionization
parameter is
\begin{equation}
\xi=\frac{4\pi m_{\rm p} c^2 h_{\rm disk} h \eta_{\rm x} (-u_r) r}{(r^2+h^2)^{3/2}}\left(\frac{1-2Gm/c^2h}{1-2Gm/c^2r}\right)^{3+\alpha},
\end{equation}
where we have used dimensionful quantities to allow numerical evaluation.
We shall assume that $h_{\rm disk}/r=0.1$ in all subsequent calculations.

Figure~3 shows the behavior of $\xi(r)$ for $h=6m$ and a
variety of efficiencies $\eta_{\rm x}\le 0.06$.  Note that $\eta_{\rm
x}=\eta=0.06$ is an upper limit to the X-ray efficiency set by the total
amount of accretion power extractable around a Schwarzschild black hole.
It can been seen that for physically reasonable X-ray efficiencies
$\eta\sim 0.01$, much of the region $r<r_{\rm ms}$ is ionized (in the sense
described in the next paragraph).  Note that,
because of our free-fall approximation for this region of the disk, {\it the
ionization is independent of the mass accretion rate and accretion
disk viscosity law.}

Fluorescent iron line emission for various ionization parameters has been
investigated by Matt, Fabian \& Ross (1993, 1996).  They come to the
following conclusions.  For small ionization parameters ($\xi\approxlt
100\erg\cm\ps$), we have the standard cold fluorescent line at $6.40\keV$.
For $\xi\sim 100-500\erg\cm\ps$, resonance scattering followed by Auger
destruction severely impedes the escape of the line photons and the
equivalent width of the line falls to very low values.  For $\xi\sim
500-5000\erg\cm\ps$, most iron ions no longer possess L-shell electrons and
so the Auger destruction mechanism cannot operate.  Ionized lines result
(at $6.67\keV$ and $6.97\keV$) and are strong due to the lack of the
competitive Auger effect.  For $\xi>5000\erg\cm\ps$, all but a negligible
fraction of the iron is fully stripped and no iron fluorescence results.
See Matt, Fabian \& Ross (1996) for a more detailed description of some of
the related atomic physics issues.

We approximate this somewhat complex situation by treating the accretion
disk as possessing four zones:
\begin{enumerate}
\item $\xi<100\erg\cm\ps$ : cold iron line at $6.4\keV$.
\item $100\erg\cm\ps<\xi<500\erg\cm\ps$ : no line emission (line photons are
resonantly trapped and destroyed via the Auger process.)
\item $500\erg\cm\ps<\xi<5000\erg\cm\ps$ : hot iron line at $6.8\keV$ with
twice the effective fluorescent yield when compared with the cold line.
\item $\xi>5000\erg\cm\ps$ : no line emission (total ionization)
\end{enumerate}
This approximation will allow us to address, in at least a
semi-quantatitive way, the effects of ionization on the fluorescent
emission from $r<r_{\rm ms}$.  However, it must be noted that we have
neglected the following two aspects.  First, Matt,
Fabian \& Ross (1996) have shown that in highly ionized disks there is a
substantial component of the line flux (in {\it addition} to that
approximated here) that suffers multiple Compton scatterings.  This line
flux emerges as a very broadened spectral feature, and so is probably
(observationally) indistinguishable from the underlying continuum.
Secondly, the Auger destruction mechanism may be less effective in the
region $r<r_{\rm ms}$ due to the very large velocity gradients that exist
in that region.   A detailed exploration of these issues is beyond the scope
of the present work.

\section{Computations of fluorescent lines}

In this section, we shall briefly describe our method for computing
emission line profiles from the accretion disk and some results.

\subsection{Computational Method}

We suppose that an observer views the accretion disk at some inclination
$i$ (with $i=0$ corresponding to a face-on disk) from a very large
distance.  In practice we place the observer at $r=1000m$.  We then
integrate photon paths from points on the image plane to the accretion disk
using the four constants of motion: energy, azimuthal angular momentum $l$,
photon rest mass (i.e. $p^\mu p_\mu=0$), and the Schwarzschild form of the
Carter constant
\begin{equation}
q=r^4\dot{\theta}^2+l^2\cot^2\theta.
\end{equation}
The integration is stopped at $\theta=\pi/2$ or $r=2.1m$, whichever
condition is met first.  Those paths that are terminated via the latter
condition are deemed to have entered the event horizon and the corresponding
point on the image plane is assigned an infinite (in practice, very large)
redshift.  However, those paths that are terminated via the $\theta=\pi/2$
condition are deemed to have struck the disk.  For those paths, we
calculate the overall redshift of photons that are emitted by the disk
material at that point of the disk and propagate to the corresponding point
of the image plane.  This redshift is given by
\begin{equation}
1+z=\frac{E}{(u^\mu p_\mu)_{\rm disk}},
\end{equation}
where $E$ is the (conserved) energy of the photon at infinity.  By
considering all points on the image plane, we can build up a map of the
redshift field on the image plane.

To calculate the profile of the iron K$\alpha$ feature, we use the
following procedure.  For a given point on the image plane, we determine
the value of $r$ for the corresponding point on the accretion disk.  For
given X-ray source parameters (i.e., $h$ and $\eta_{\rm x}$), this allows us
to determine the ionization parameter $\xi(r)$ and illumination $g(r)$ at
that point on the disk.  Using our simple prescription for line emission
from ionized material, we determine whether this part of the disk emits no
line, a cold line or an ionized line.  We assume that any iron line
emission is isotropic in the comoving frame of the disk material.  More
sophisticated treatments go beyond this assumption in order to deal with
limb-darkening effects, but we shall not be concerned with these
complications.  The neglect of limb-darkening will only become important
for high-inclination disks.  For a given point on the image plane, the
observed intensity of the line is given by
\begin{equation}
I_{\rm obs}=\frac{I_0Yg(r)}{(1+z)^4}
\end{equation}
where $I_0$ is some overall normalization, $Y$ is the effective fluorescent
yield, and we have utilized the phase space invariant $I_\nu/\nu^3$.  Note
that the line flux (which is a frequency-integrated quantity) depends upon
$(1+z)^4$, and not $(1+z)^3$, due to the transformation of the frequency
element (Dabrowski et al. 1997).  

Thus, for a given illumination pattern $g(r)$, we can build up an iron line
intensity map of the disk on the image plane.  This can be combined with
the redshift map (taking into account the different energies of the cold
and ionized iron lines) in order to produce an overall line profile.

\subsection{Results}

Given the assumptions stated above, the observed line profile is a function
of the inclination $i$, the illumination law $g(r)$, and the ionization
structure $\xi(r)$.  For the point source geometry of Section~2.3, the form
of the illumination law is determined purely by $h$ and the ionization
structure is determined by $h$ and $\eta_{\rm x}$ according to eq. (19).
We will organize our discussion by addressing systems with various
efficiencies.

\subsubsection{Very inefficient sources}

\begin{figure*}
\hbox{
\psfig{figure=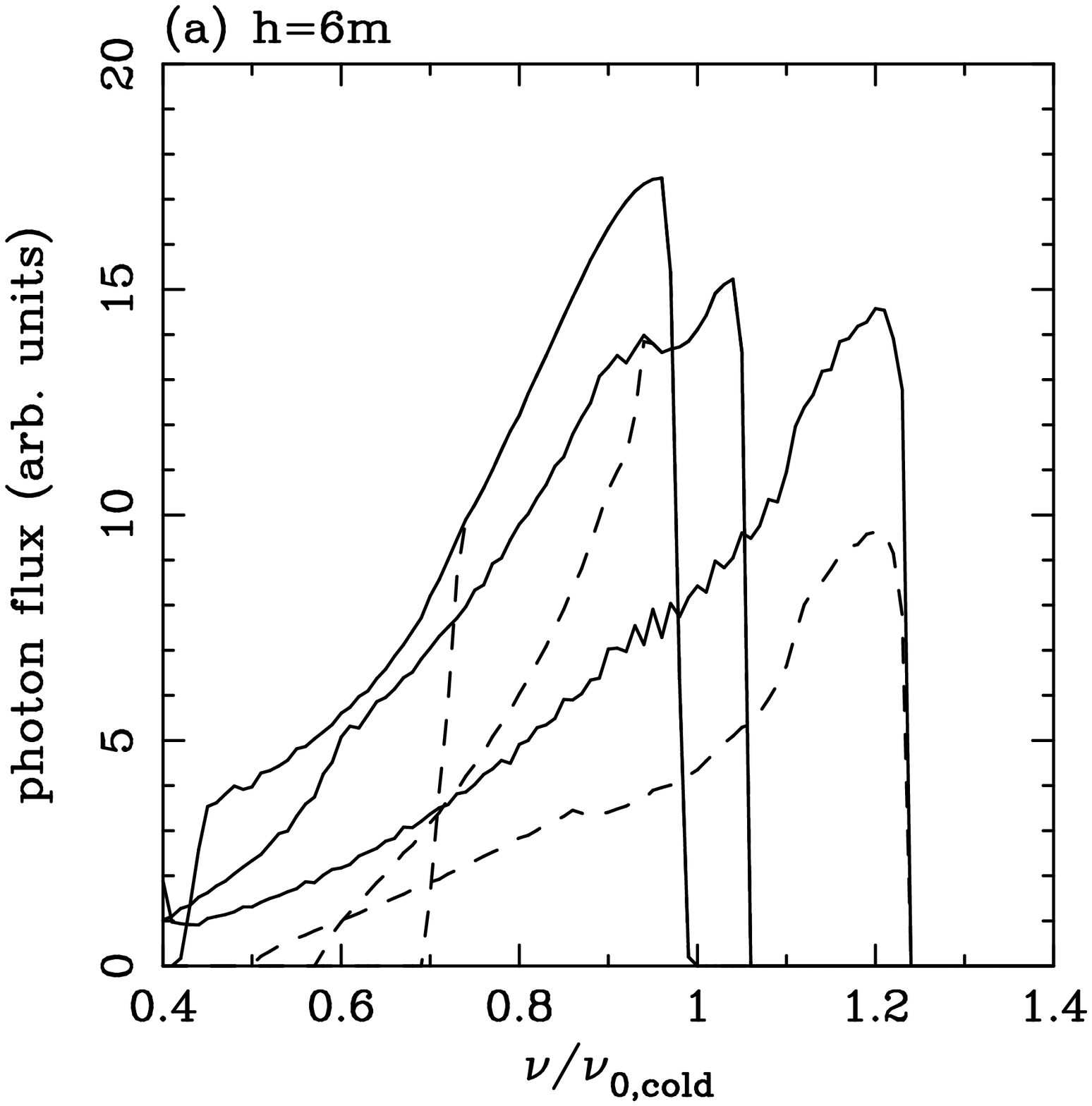,width=0.6\textwidth,angle=270}
\hspace{-2cm}
\psfig{figure=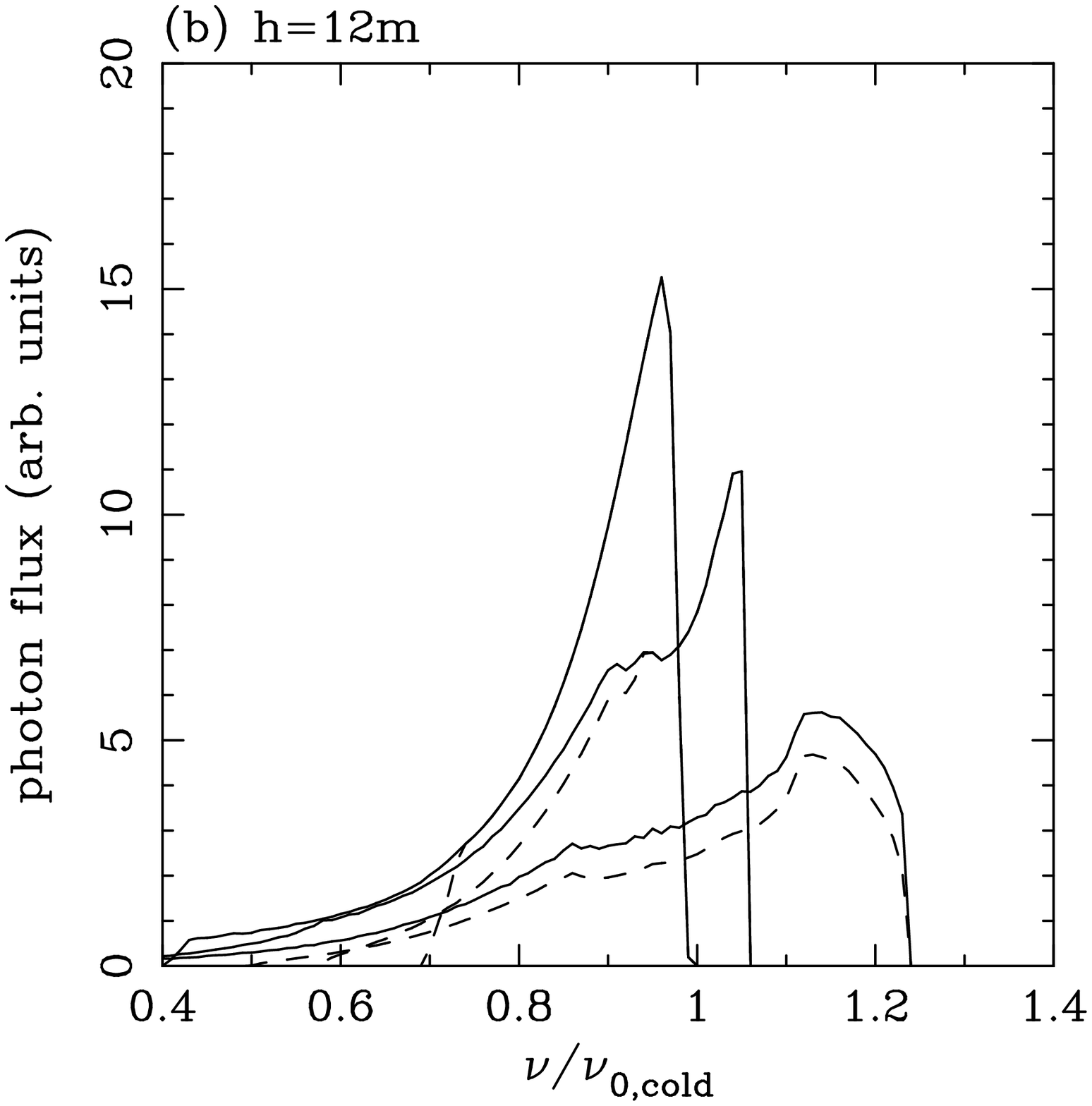,width=0.6\textwidth,angle=270}
}
\caption{Iron line profiles from very inefficient sources ($\eta_{\rm
x}=10^{-5}$) for $i=0\degmark$, $i=30\degmark$ and $i=60\degmark$ (line peak
moves to higher frequency for increasing inclination).  The dashed curves
show the line profile resulting from material that is outside the innermost
stable orbit, whereas the solid curves show the total line profile
(i.e., including the region $r<r_{\rm ms}$).  Results are shown for two
values of the source height: (a) $h=6m$, and (b) $h=12m$.  The rest-frame
frequency of the cold iron K$\alpha$ line (6.4\,keV) is denoted by
$\nu_{\rm 0,cold}$.}
\end{figure*}

\noindent Suppose that the X-ray source is so inefficient that most of the 
region $r<r_{\rm ms}$ produces a cold iron line.  For $h=6m$, this
translates to a limit on the efficiency of $\eta_{\rm x}\approxlt 10^{-5}$
(see Fig.~3).  Low X-ray efficiencies may be obtained if the bulk of the
accretion energy is radiated at lower frequencies (e.g., in the Big Blue
Bump), or is directly converted into bulk kinetic energy of a wind or jet.

Figure~4 shows the resulting line profiles for the cases $h=6m$ and
$h=12m$.  Comparing the full line profiles (solid lines) with the
contribution just from the region outside the innermost stable orbit
($r>r_{\rm ms}$; dashed lines), we can say the following.  First, the
region within the innermost stable orbit is a major contributor to the
observed iron line for $h=6m$, but not for $h=12m$.  This is simply because
the former illumination law is more centrally concentrated than the latter.
Secondly, in low-inclination sources, the line emission from the region
$r<r_{\rm ms}$ emerges in the red-wing of the observed line due to the
strong gravitational redshifts experienced by these photons.  In
high-inclination sources this emission can also affect the blue peak of the
line due to the large line-of-sight velocities that are attained by the
material in this inner region.  However, even in these high-inclination
systems, the overall effect is to significantly enhance the red-wing
relative to the blue peak.

\subsubsection{Very efficient sources}

\begin{figure*}
\hbox{
\psfig{figure=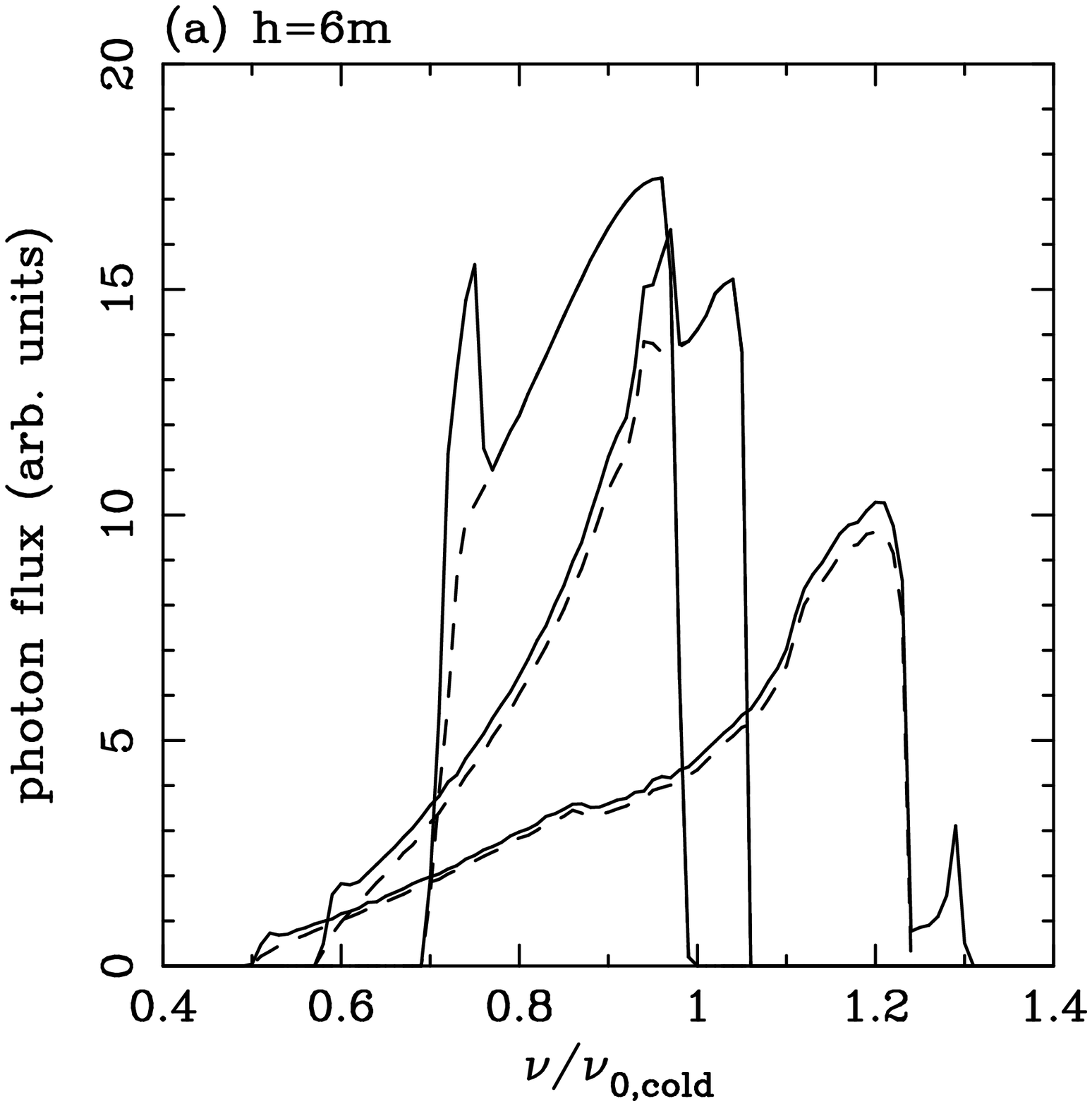,width=0.6\textwidth,angle=270}
\hspace{-2cm}
\psfig{figure=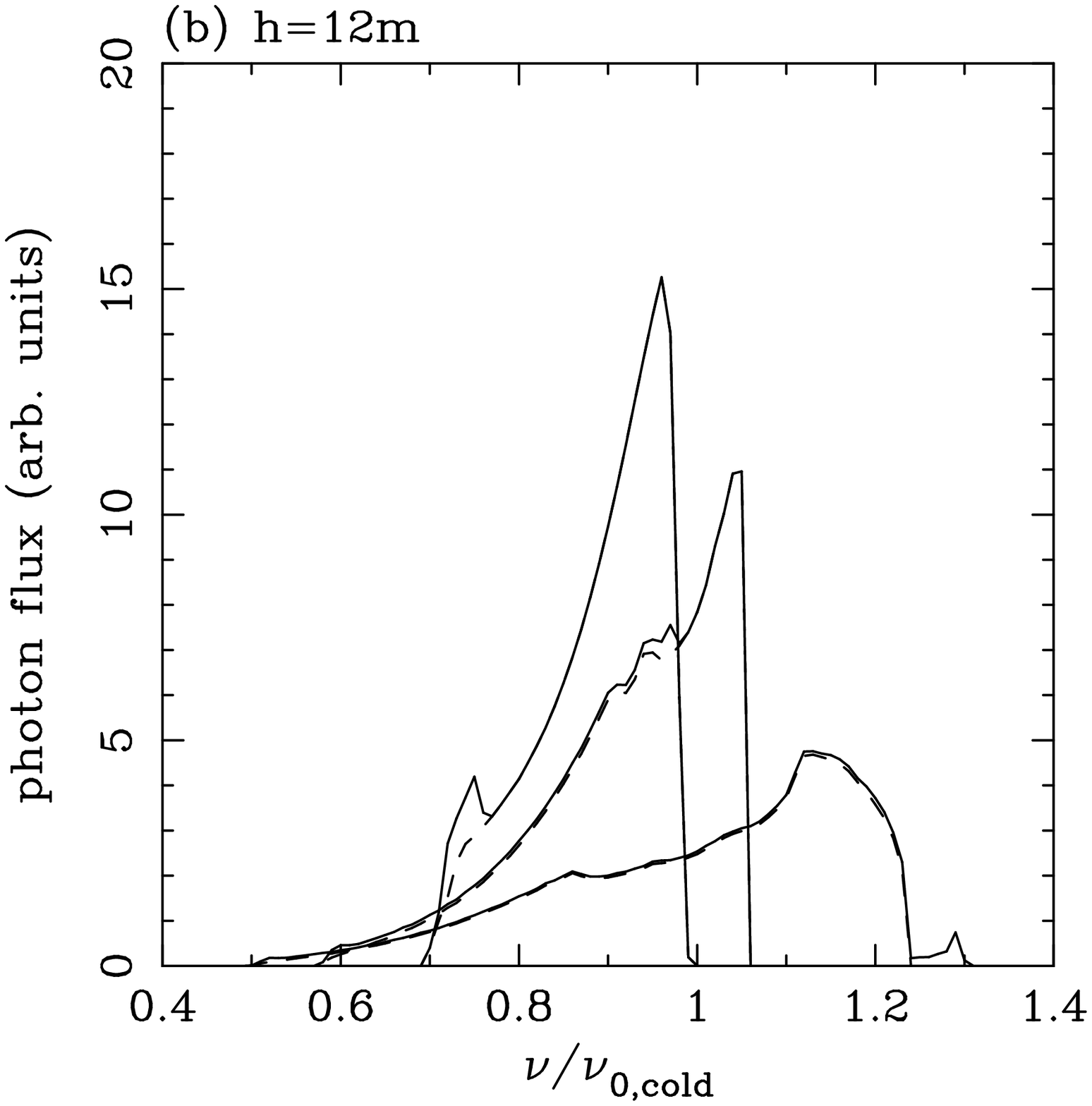,width=0.6\textwidth,angle=270}
}
\caption{Iron line profiles from very efficient sources ($\eta_{\rm
x}=0.06$) for $i=0\degmark$, $i=30\degmark$ and $i=60\degmark$ (line peak
moves to higher frequency for increasing inclination).  The dashed curves
show the line profile resulting from material that is outside the innermost
stable orbit, whereas the solid curves show the total line profile
(i.e., including the region $r<r_{\rm ms}$).  Results are shown for two
values of the source height: (a) $h=6m$, and (b) $h=12m$.  The rest-frame
frequency of the cold iron K$\alpha$ line (6.4\,keV) is denoted by
$\nu_{\rm 0,cold}$.}
\end{figure*}

\noindent Now consider a maximally-efficient X-ray source, $\eta_{\rm
x}=0.06$ (i.e., all extractable accretion power is channeled into the
X-ray source).   As shown in Section 2.4, a source close to the disk with
such an efficiency will ionize most of the region within the innermost
stable orbit.    However, there will be a (narrow) annulus of this region
near $r=r_{\rm ms}$ that will be sufficiently dense, and thus have a
sufficiently low ionization parameter, so as to produce iron fluorescence
(from helium and hydrogen-like iron).

Figure~5 shows the line profiles from such efficient sources.  It can be
seen that, for face-on disks, the contribution from the region $r<r_{\rm
ms}$ is almost monochromatic due to the fact that only a narrow annulus of
this region is contributing to the fluorescence.  The energy of this narrow
spectral feature simply corresponds to the redshifted line of ionized iron.
For finite inclinations, the ionized fluorescing annulus produces a
classical double-peaked line shape which can be seen superposed on the
`cold' broad line from the region outside the innermost stable orbit.

\section{Discussion}

\subsection{The case of MCG$-$6-30-15}

The Seyfert 1 galaxy MCG$-$6-30-15 displays one of the best studied
accretion disk iron-lines, and so provides a good laboratory in which to
examine our model.  The time-averaged line profile is well explained as
originating from the region $r\approx 6m-20m$ of an accretion disk around
{\it either} a Schwarzschild or a Kerr black hole (Tanaka et al. 1995).
However, I96 found the line profile to be variable.  In
the most dramatic event, the red-wing of the line becomes extremely broad
and strong.  During this event, line emission can be discerned down to at
least $4\keV$ and the equivalent width of the line increases by a factor of
3.  As described in the Introduction, this led to the suggestion that
MCG$-$6-30-15 harbors a near-extremal Kerr hole and that the
line-broadening event corresponds to a shift in the X-ray illumination that
suddenly favored the innermost regions of the disk.  Dabrowski et
al. (1997) examined the whole family of Kerr geometries and found that the
very-broad state of this iron line (and the assumption of no fluorescence
from within the innermost stable orbit) requires $a>0.94$.

If vindicated, this is an extremely important result.  As well as being the
first {\it observational} evidence of a rapidly rotating black hole, it has
significant implications for the nature of the radio-quiet/radio-loud
dichotomy observed in accreting black hole systems.  Motivated by the
discovery that hydromagnetic processes can efficiently extract the
spin-energy of a rapidly rotating black hole (Blandford \& Znajek 1977), it
has long been suggested that the spin of the central black hole is the key
physical quantity that determines whether the system can produce
collimated, relativistic radio jets, which lead to a radio-loud
classification (Rees et al. 1982; Blandford 1990; Wilson \& Colbert 1995).
However, MCG$-$6-30-15 appears to represent a paradigm-destroying
counter-example: it is a radio-quiet Seyfert nucleus which seems to possess
a near-extremal Kerr black hole.  Given its importance, the Kerr
interpretation for MCG$-$6-30-15 must be critically examined and tested
against alternative models.

\begin{figure*}
\hspace{0cm}
\psfig{figure=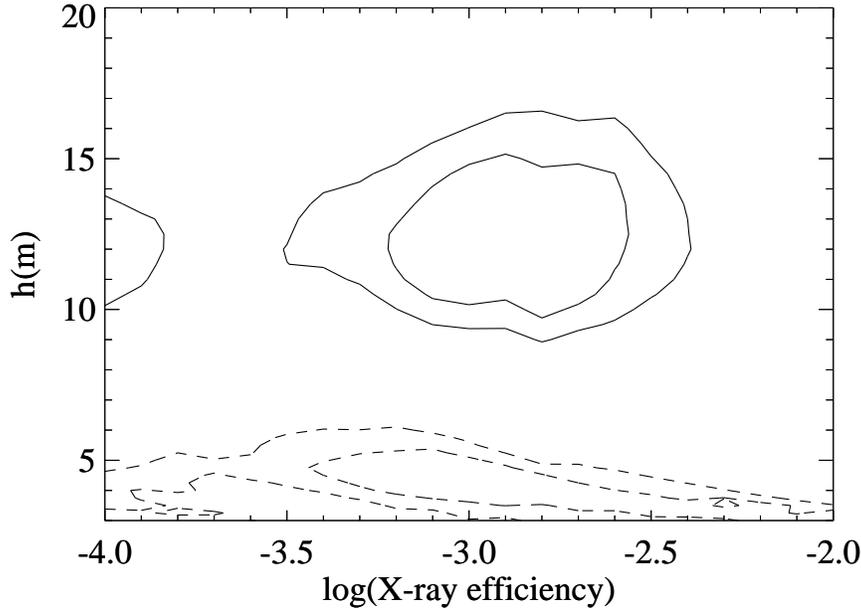,width=0.8\textwidth}
\caption{Confidence contours for the disk-line model parameters 
(including fluorescence from $r<r_{\rm ms}$) upon comparison with the data
for MCG$-$6-30-15.  The solid contours show the fit to the time-averaged
line (from Tanaka et al. 1995) whereas the dashed contours show the fit to
the line in its very-broad state (from I96).  1-$\sigma$
and 90 per cent contours are shown.}
\end{figure*}

\begin{figure*}
\hbox{
\psfig{figure=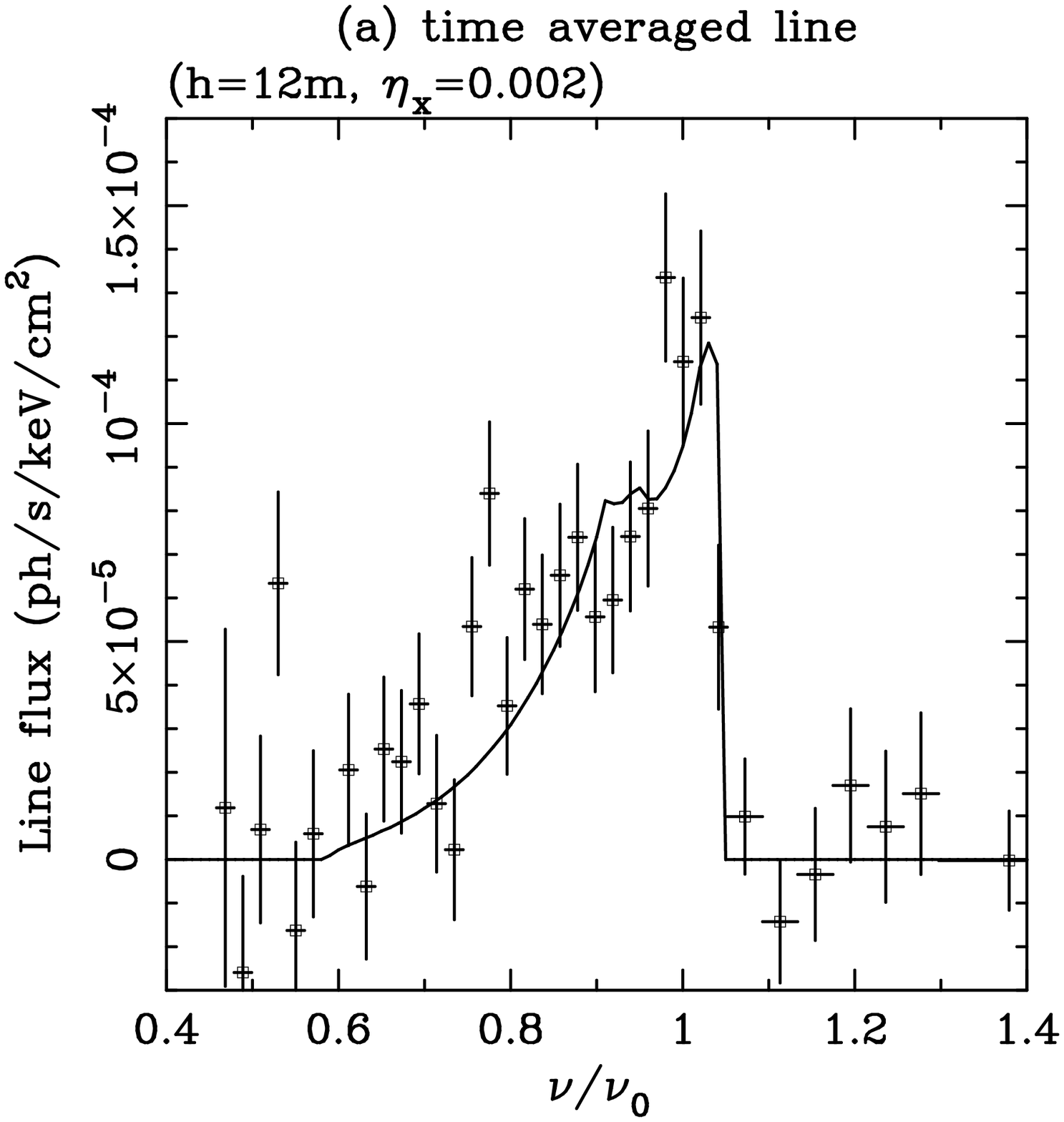,width=0.6\textwidth,angle=270}
\hspace{-2cm}
\psfig{figure=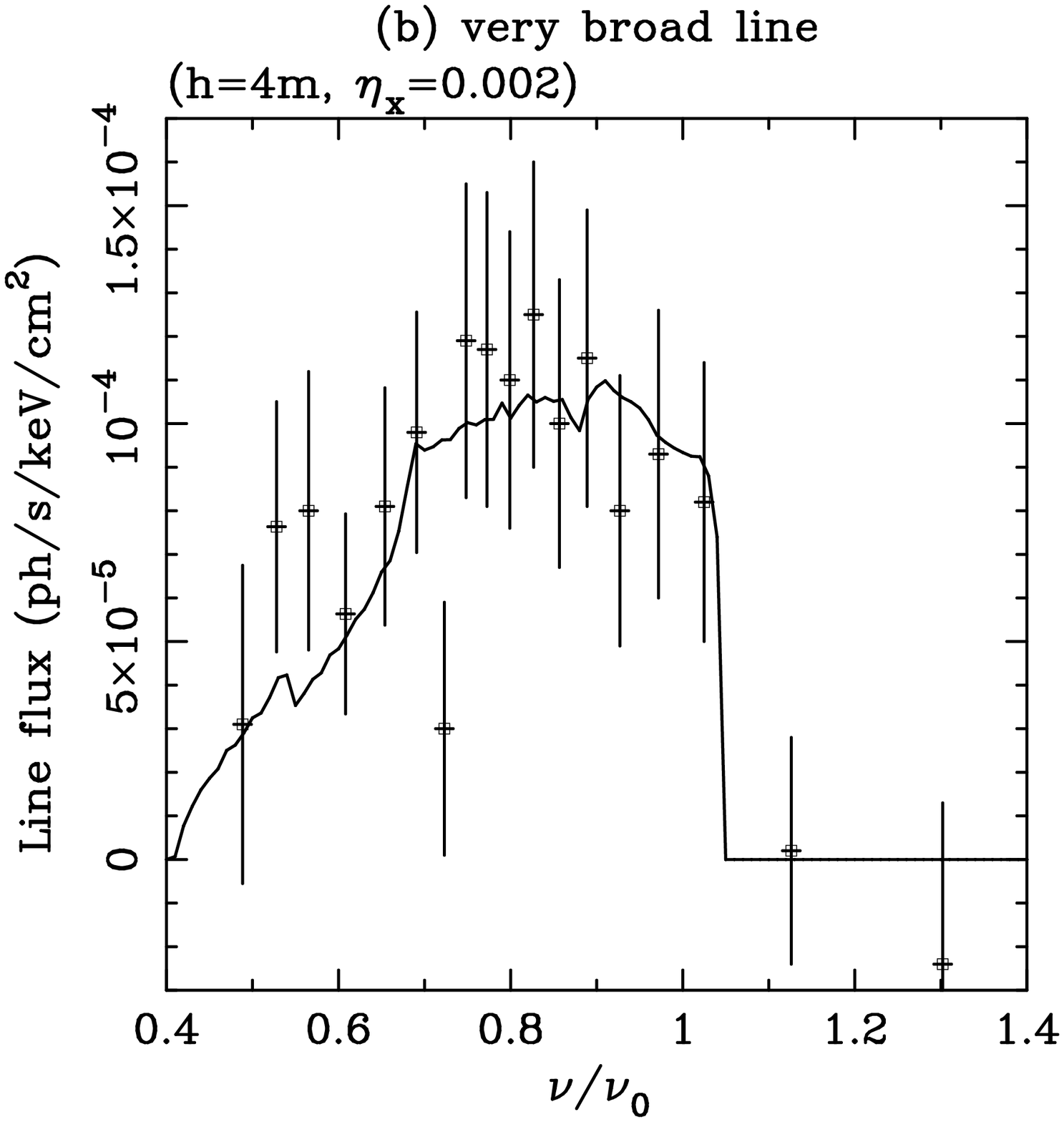,width=0.6\textwidth,angle=270}
}
\caption{Best fit disk-line models (including fluorescence from $r<r_{\rm
ms}$) for (a) the time-averaged line of MCG$-$6-30-15, and (b) the
very-broad state of that line.}
\end{figure*}

Here, we ask the following question: can the extremely broad state of the
iron line in MCG$-$6-30-15 be reconciled with a Schwarzschild (or slowing
rotating Kerr) black hole, if we also allow for the possibility of line
emission from within the innermost stable orbit?  We address this by
comparing (in a $\chi^2$ sense) the time-averaged line profile (from Tanaka
et al. 1995), and the very-broad line profile (from I96), with our
calculated line profiles for various values of $h$ and $\eta_{\rm x}$.  In
performing these fits, the inclination has been fixed at $i=28\degmark$.
The inclination is determined in a robust manner by the energy at which the
blue side of the line truncates.  Figure~6 shows the resulting confidence
contours on the ($\eta_{\rm x}$, $h$) plane, and Fig.~7 displays the best
fit models overlaid on the respective datasets.  It is seen that the
very-broad line in MCG$-$6-30-15 can, indeed, be understood in the context
of Schwarzschild geometry.  As in the Kerr geometry scenario, a dramatic
change in the X-ray illumination law is required in order to explain the
transition from one line profile to the other.  Within the context of our
point source geometry, the principal change is that the source height
decreases from $h_1\approx 12m$ down to $h_2\approx 4m$.  The profiles are
consistent with the X-ray efficiency remaining constant at $\eta_{\rm
x}\approx 0.002$ (i.e., 3 per cent of the total extractable accretion energy
is channelled into the X-ray flux of this source).

Although the point source geometry adopted here is somewhat artificial, it
is interesting to explore this model further.  Suppose that the line
variability is caused by the source moving from $h_1$ to $h_2$ at constant
$\eta_{\rm x}$.  Assuming a black hole mass of $M\sim 10^7\Msun$, the
time-scale of the line variability ($\sim 10\,000\s$) is only a few times the
dynamical timescale at the innermost stable orbit, and is very much less
than the viscous timescale.  Thus, we would not expect the mass accretion
rate $\dot{m}$ to vary on this timescale and, for fixed $\eta_{\rm x}$, the
intrinsic luminosity of the X-ray source will be constant.  The {\it
observed} luminosity will be diminished by the increasing gravitational
effects as the source moves close to the hole.  The factor by which the
observed luminosity changes is approximately
\begin{equation}
\left(\frac{1-2m/h_2}{1-2m/h_1}\right)^{(3+\alpha)/2}\approx 0.36.
\end{equation}
This compares very well with the factor of 2--3 decrease in the observed
continuum level that accompanies the change in line profile.  Continuing
with this hypothesis, when the source moves closer to the hole, an
increasing fraction of the emitted luminosity is gravitationally focused
toward the disk plane.  All other things being equal, this will increase
the equivalent width of the iron line.  From our line computations, and
taking into account the fall of the observed continuum, we would expect the
equivalent width of the line to increase by a factor of $\sim 3.4$.  This
compares very well with the factor of $\sim 3$ increase in the observed
equivalent width.

A cautionary note is in order.  A backscattered continuum will accompany the
observed iron line emission, and this continuum will have iron K-shell
photoelectric edges imprinted into it (e.g., see Ross \& Fabian 1993 for a
calculation of the backscattered continuum for various ionization states of
the disk).  The large spread in redshifts that contribute to these
reprocessing features results in a smearing of these edges into broad
troughs (Ross, Fabian \& Brandt 1996).  Thus, the net continuum that
underlies the I96 iron line will, at some level, possess these broad
troughs.  Since the line profile shown in Fig.~7b (from I96) has been
derived from the data by subtracting a continuum with no such smeared
edges, the {\it true} emission line profile may differ to some extent from
that used in this work.  Further modeling, and higher
signal-to-noise datasets (with a better constrained high-energy continuum
shape) are required to address this complication.

\subsection{The transparency of the region $r<r_{\rm ms}$}

Appreciable line fluorescence can occur only in regions where $\tau_{\rm
e}\approxgt {\rm few}$.  Provided $L\approxgt 0.1~L_{\rm Edd}$, this
condition is satisfied essentially all the way down to the event horizon
(assuming $\eta=0.06$).  However, for systems with smaller Eddington
ratios, the innermost regions of the disk can be Thomson thin.  By scaling
the curve in Fig.~1, it can be seen that for $L\sim 0.01~L_{\rm Edd}$, the
region within $r=5m$ will be not be able to produce significant
fluorescence.  For significantly lower Eddington ratios, virtually no iron
line photons will be produced within the innermost stable orbit.  Note that
for high Eddington ratios, $L\approxgt 0.3~L_{\rm Edd}$, we expect the
central regions of the disk to become geometrically thick, thereby
complicating the geometry beyond that assumed in our model.

If fluorescence from $r<r_{\rm ms}$ is important in real systems, the
dependence of $\tau_{\rm e}$ on Eddington ratio implies the following
observationally-testable prediction: as one considers systems with smaller
Eddington ratios, fluorescence from within $r<r_{\rm ms}$ will become
progressively less important and the red-wing of the iron line will become
less broad.  This effect might reveal itself as a (positive) correlation
between source luminosity and the breadth of the iron line, and an
anticorrelation between the source luminosity and centroid energy of the line.
However, different black hole masses from object to object will induce a
large scatter in any correlations.  Thus, large samples, covering a
wide range of luminosities, will be required to search for this
correlation.  Alternatively, in sources where we have an independent
measure of the Eddington ratio, this issue could be confronted directly.
An example of such a source is NGC~4258 where studies of the MASER disk
imply that $L\sim 10^{-3}L_{\rm Edd}$ (Neufeld \& Maloney 1995).  Both of
these tests require the use of future, high-throughput X-ray missions, such
as {\it XMM}.

\subsection{Distinguishing Kerr and Schwarzschild geometries}

On the basis of the study presented in this paper, it is apparent that the
observation of an extensive red-wing on an iron line is not sufficient
evidence, by itself, that the black hole is rapidly rotating.  Clearly,
further constraints are required in order to distinguish, observationally,
a rapidly rotating black hole from a slowly rotating hole.

The largest uncertainty is the nature (both geometry and efficiency) of the
X-ray source.  The geometry of the source is very difficult to address
observationally with current data.  However, given various assumptions, it
is possible to set a lower limit to the X-ray efficiency by studying the
most rapid continuum variations (Fabian 1979).  The principal assumptions
that enter into this argument are that the continuum variability is not
caused by relativistic beaming, that the X-ray source is spherical, and
that the X-ray energy is deposited locally by the accreting matter.
Reynolds et al. (1995) have applied the Fabian efficiency limit to
MCG$-$6-30-15 and find $\eta_{\rm x}\approxgt 0.06$.  Such a high
efficiency alone may imply accretion onto a Kerr hole.  Also, a source of
this efficiency within the scenario considered here would almost certainly
ionize the disk within the innermost stable orbit, thereby lending strength
to the Kerr hole interpretation for the MCG$-$6-30-15 iron line.  However,
it is unclear how the Fabian efficiency limit is affected by a relaxation
of these assumptions -- in practice the source may not be spherical, and
the X-ray energy is almost certainly not deposited locally in the accretion
flow but, rather, is deposited in a disk-corona.  More theoretical work is
required in order to determine whether the efficiency limit can be relaxed
sufficiently to permit fluorescence from within the innermost stable orbit.

Multiwavelength studies are also important in constraining the X-ray
efficiency of a given AGN.   After accounting for the various reprocessing
mechanisms, Reynolds et al. (1997) argue that $\sim 20-50$ per cent of the
primary radiant energy in MCG$-$6-30-15 is released in the X-ray regime.
Assuming that all of the accretion energy is radiated, this argues that
\begin{equation}
\eta_{\rm x}\approx (0.2-0.5)\eta\sim 0.01-0.03,
\end{equation}
where we have taken the overall efficiency to be the Schwarzschild value,
$\eta=0.06$.  The uncertain reddening correction is the main source of this
large spread in inferred X-ray fractions.  Thus, the X-ray efficiency
required by our model (see Fig.~6) is only barely consistent with the lower
end of this range.  However, if some fraction of the accretion energy is
channeled into a non-radiant form (such as the kinetic energy of a disk
wind), the multi-waveband studies become more consistent with our model.

Eventually, we will be able to map the geometry of the illuminating X-ray
source.  Observations with high-throughput X-ray observatories will be able
to map the response time of the line to observed continuum flares.  If the
response is almost immediate, we can infer that the flaring X-ray sources
are located just above the disk surface.  Within this geometry, it would be
difficult to illuminate the region within the innermost stable orbit, and
little fluorescence would result from this region.  This would strengthen
the near-extremal Kerr hole interpretation for those objects possessing
very broad iron lines of the type seen by I96.  On the other hand, if a
response lag corresponding to a few gravitational radii or more is seen, we
can argue for an X-ray source that is somewhat displaced from the disk.
This would allow significant fluorescence to be excited from within the
innermost stable orbit, thereby lending evidence to the picture explored in
this work.  For a central black hole mass of $M\sim 10^7\Msun$, one
gravitational radius has a light crossing time of $\sim 50\s$.  We need to
be able to determine line strengths on this timescale in order to make this
observational test feasible.  Even for the brightest objects, this would
require an instrument with an effective area of $\sim 5000\cmsq$ at
$6\keV$.  This is just within the specified capabilities of the {\it High
Throughput X-ray Spectroscopy Mission} ({\it HTXS}).

The most straightforward, but observationally intensive, means of
determining whether a given black hole is rotating rapidly is to perform
high-throughput, high-resolution X-ray spectroscopy of the iron line.  Matt
et al. (1992, 1993) have shown that strong-gravitational effects produce
fine-structure in the line profile.  These features are especially
pronounced in high-inclination sources, and can be used as a probe of
strong light-bending and frame dragging.  In principle, fitting detailed
models to high-quality data will allow the inclination, Kerr parameter, and
illumination law to be constrained.

\section{Conclusions}

We have explored the possibility that iron fluorescence from {\it inside}
the innermost stable orbit of a black hole accretion disk may be
observable.  For concreteness, we have examined the case of a (simplified)
accretion disk around a Schwarzschild hole which is illuminated by a point
X-ray source located on the rotation axis of the disk at some height above
the disk plane.  The principal assumptions that enter into the accretion
disk structure are that it is thin ($h_{\rm disk}/r=0.1$) and in free-fall
within the innermost stable orbit.  While this is an idealized case, it is
likely to be qualitatively applicable to accretion disks around
slowly-rotating black holes that are illuminated by some
geometrically-thick, and centrally concentrated, X-ray source.

For mass accretion rates that are thought to be relevant to some Seyfert
nuclei ($L\sim 0.1~L_{\rm Edd}$), it is found that the disk remains
optically thick to electron scattering right down to the event horizon.
Thus, this region can produce appreciable iron fluorescence provided that
two conditions are satisfied.  Firstly, the X-ray illumination must be
centrally-concentrated in order to produce an observable fluorescent
signature from this region.  The intense gravitational focusing of X-rays
onto these central regions of the disk can readily produce the required,
centrally-condensed, X-ray illumination law, provided that the source is
located relatively close to the disk plane ($h\approxlt 10m$).  Secondly,
fluorescence is possible only if the material in this region is {\it not}
fully ionized.  Since the ionization parameter is essentially the ratio of
the X-ray flux to the density, it depends upon both the geometry of the
source and the efficiency with which accretion power is transformed into
X-ray power.

We present calculations of line profiles for various values of inclination
$i$, source height $h$, and X-ray efficiency $\eta_{\rm x}$.  In the limit
of very-inefficient sources, the entire region $r<r_{\rm ms}$ produces a
`cold' iron line at 6.4\,keV.  Due to the enormous gravitational redshifts
experienced by these line photons, the effect is to enhance the
red wing of the observed iron line.  This effect is stronger for larger
inclinations since the relativistic motion of the material in this region
tends to beam the emission into the disk plane.  In the opposite limit of
very-efficient sources ($\eta_{\rm x}=\eta=0.06$), only a narrow annulus
within the innermost stable orbit is sufficiently combined to produce a
(hot) iron line.  For face-on sources, the fact that this extra line
contribution originates from a narrow annulus leads to a narrow spectral
feature at $4.6\keV$.  For finite inclinations, the emission from this
annulus has a classical double peaked line profile.

In the light of these calculations, we re-examine the iron line variability
seen in the Seyfert 1 galaxy MCG$-$6-30-15 by I96.  During certain periods,
this source is seen to possess an iron line with an extremely broad red
wing (line emission can be discerned down to 4\,keV and below) and a very
high equivalent width.  If one assumes no fluorescence from within the
innermost stable orbit, one is led to conclude the existence of a
near-extremal Kerr black hole in this Seyfert 1 nucleus (Iwasawa et
al. 1996; Dabrowski et al. 1997).  This result would appear to argue
against the black hole spin paradigm for the radio-loud/radio-quiet
dichotomy.  However, fluorescence from the region $r<r_{\rm ms}$ may also
be relevant to this observation.  Within the framework of our model (with
Schwarzschild geometry), the line profile variability is well explained as
resulting from a lowering of the X-ray source from $h\approx 12m$ to
$h\approx 4m$, with a constant source efficiency $\eta_{\rm x}$.  The
resulting increase in the gravitational focusing of these X-rays toward
the disk-plane can explain the observed changes in both the continuum flux
and the iron line equivalent width, although we note that we have not
accounted for the effects of the accompanying iron edge.  We discuss ways
of distinguishing between the rapidly rotating hole model and slowly
rotating hole model using observational constraints on source geometry and
efficiency.

\section*{Acknowledgments}

We are very grateful to Andy Fabian and Martin Rees for their useful
comments throughout the course of this work.  We thank Kazushi Iwasawa for
kindly providing the data on MCG$-$6-30-15.  This work has been supported
by the National Science Foundation under grant AST9529175.


\begin{thebibliography}{}
\bibitem{} Abramowicz M.~A., Kato S., 1989, ApJ, 336, 304
\bibitem{} Blandford R.~D., 1990, in Active Galactic Nuclei, ed
T.J.-L.Courvoisier \& M.Mayor (Saas-Fee Advanced Course 20)
(Berlin:Springer), 161
\bibitem{} Blandford R.~D., Znajek R., 1977, MNRAS, 179, 433
\bibitem{} Chen X., Taam R.~E., 1993, ApJ, 412, 254
\bibitem{} Dabrowski Y., Fabian A.~C., Iwasawa K., Lasenby A.~N., Reynolds
C.~S., 1997, in press
\bibitem{} Fabian A.~C., 1979, Proc.~Roy.~Soc.~, A366, 449
\bibitem{} Fabian A.~C., Rees M.~J., Stellar L., White N.~E., 1989, MNRAS,
238, 729
\bibitem{} Fabian A.~C. et al. 1995, MNRAS, 277, L11
\bibitem{} George I.~M., Fabian A.~C., 1991, MNRAS, 249, 352
\bibitem{} Iwasawa K. et al., 1996, MNRAS, 282, 1038 (I96)
\bibitem{} Laor A., 1991, ApJ, 376, 90
\bibitem{} Martocchia A., Matt G., 1996, MNRAS, 282, L53
\bibitem{} Matt G., Fabian A.~C., Ross R.~R., 1993, MNRAS, 262, 179
\bibitem{} Matt G., Fabian A.~C., Ross R.~R., 1996, MNRAS, 278, 1111
\bibitem{} Matt G., Perola G.~C., Piro L., 1991, A\&A, 247, 25
\bibitem{} Matt G., Perola G.~C., Piro L., 1992, A\&A, 257, 63
\bibitem{} Matt G., Perola G.~C., Stellar L., 1993, A\&A., 267, 643
\bibitem{} Misner C.~W., Thorne K.~S., Wheeler J.~A., 1973, Gravitation,
Freeman, San Francisco.
\bibitem{} Muchotrzeb B., Pacynski B., 1982, Acta Astr., 32, 1
\bibitem{} Mushotzky R.~F., Fabian A.~C., Iwasawa K., Kunieda H., Matsuoka
M., Nandra K., Tanaka Y., 1995, MNRAS, 272, L9
\bibitem{} Nandra K., George I.~M., Mushotzky R.~F., Turner T.~J., Yaqoob
T., 1997, ApJ, 477, 602
\bibitem{} Nandra K. et al., 1995, MNRAS, 276, 1
\bibitem{} Nandra K., George I.~M., Turner T.~J., Fukazawa Y., 1996, ApJ,
464, 165
\bibitem{} Neufeld D.~A., Maloney P.~R., 1995, ApJ, 447, L17
\bibitem{} Pounds K.~A., Nandra K., Stewart G.~C., George I.~M., Fabian
A.~C., 1990, Nat, 344, 132
\bibitem{} Rees M.~J., Begelman M.~C., Blandford R.~D., Phinney E.~S.,
1982, Nat, 295, 17
\bibitem{} Reynolds C.~S., 1997, MNRAS, 286, 513
\bibitem{} Reynolds C.~S., Fabian A.~C., submitted
\bibitem{} Reynolds C.~S., Fabian A.~C., Inoue H., 1995, MNRAS, 276, 1311
\bibitem{} Reynolds C.~S., Fabian A.~C., Nandra K., Inoue H., Kunieda H.,
Iwasawa K., 1995, MNRAS, 277, 901
\bibitem{} Reynolds C.~S., Ward M.~J., Fabian A.~C., Celotti A., 1997,
submitted 
\bibitem{} Ross R.~R., Fabian A.~C., 1993, MNRAS, 261, 74
\bibitem{} Ross R.~R., Fabian A.~C., Brandt W.~N., 1996, MNRAS, 278, 1082
\bibitem{} Tanaka Y. et al.,  1995, Nat, 375, 659
\bibitem{} Wilson A.~S., Colbert E.~J.~M., 1995, ApJ, 438, 62
\end{thebibliography}
\end{document}